\documentclass[aps,prl,floatfix,preprint]{revtex4}

\usepackage{graphicx}
\newcommand{\lyxmathsym}[1]{\ifmmode\begingroup\def\b@ld{bold}
  \text{\ifx\math@version\b@ld\bfseries\fi#1}\endgroup\else#1\fi}

\begin{document}

\title{Beyond spontaneously broken symmetry in
Bose-Einstein condensates}
\author{W. J. Mullin$^{a}$ and F. Lalo\"{e}$^{b}$}
\affiliation{$^{a}$Department of Physics, University of
Massachusetts, Amherst,
Massachusetts 01003 USA\\
$^{b}$ Laboratoire Kastler Brossel, ENS, UPMC, CNRS; 24 rue Lhomond,
75005
Paris, France}
\email{mullin@physics.umass.edu;laloe@lkb.ens.fr}

\begin{abstract}
Spontaneous symmetry breaking (SSB) for Bose-Einstein condensates
cannot treat phase off-diagonal effects, and thus not explain Bell inequality
violations. We describe another situation that is beyond a SSB treatment: an
experiment where particles from two (possibly
macroscopic) condensate sources are used for conjugate measurements of the relative phase and
populations. Off-diagonal phase effects are characterized by a ``quantum angle'' and
observed via ``population oscillations'', signaling quantum interference of macroscopically distinct states (QIMDS).
\end{abstract}

\maketitle

If two or more Bose-Einstein condensates (BEC) merge, they produce
an interference pattern in their densities, as shown by spectacular
experiments with alkali atoms \cite{WK}. The usual explanation assumes
spontaneous symmetry breaking (SSB) of particle number conservation, where each condensate gets
a (random) classical phase and a macroscopic wave function:
\begin{equation}
\left\langle \psi_{\alpha,\beta}(\mathbf{r})\right\rangle =\sqrt{n_{\alpha,\beta}(\mathbf{r})}\;e^{i\phi_{\alpha,\beta}(\mathbf{r)}}\label{eq:BrokenSym}\end{equation}
where $n_{\alpha,\beta}(\mathbf{r)}$ and $\phi_{\alpha,\beta}$ are density and phases of condensates $\alpha$, $\beta$.
Alternatively, one can use a ``phase state'' describing two condensates with a relative phase
$\phi$ and a fixed total number of particles:
\begin{equation}
\left|\phi,N\right\rangle =\frac{1}{\sqrt{2^{N}N!}}\;(a_{\alpha}^{\dagger}+e^{i\phi}a_{\beta}^{\dagger})^{N}\left|0\right\rangle \label{eq:Phase state}\end{equation}
where $a_{\alpha}^{\dagger}$ and $a_{\beta}^{\dagger}$ create particles
in condensates $\alpha$ and $\beta$, respectively. However, one
can also consider that two condensates are more naturally described
by a double Fock state (DFS), a state of definite particle numbers,
for which the phase is completely undetermined:
\begin{equation}
\left|N_{\alpha}N_{\beta}\right\rangle =\frac{1}{\sqrt{N_{\alpha}!N_{\beta}!}}\;a_{\alpha}^{\dagger N_{\alpha}}a_{\beta}^{\dagger N_{\beta}}\left\vert \text{0}\right\rangle \label{initialstate}\end{equation}
It is found \cite{Java}-\cite{LM-1} that repeated quantum measurements
of the relative phase of two Fock states can make a well-defined value
emerge spontaneously, but with a random value. For example,
the probability of finding $M$ particles, out of a total of $N$,
at positions $\mathbf{r}_{1}\lyxmathsym{\ldots}\mathbf{r}_{M}$ where
$M\ll N$ is shown to be given by \cite{FL,LM-1}:
\begin{equation}
P(\mathbf{r}_{1},\cdots\mathbf{r}_{M})\sim\int_{-\pi}^{\pi}\frac{d\lambda}{2\pi}\prod_{i=1}^{M}\left[1+\cos(\mathbf{k}\cdot\mathbf{r}_{i}+\lambda)\right]\label{classical}
\end{equation}
Positions can be obtained one by one from this distribution; for
large enough $M$ the integrand peaks sharply \cite{MLK}
at a single value, just as a particular phase is found in the
interference measurement of Ref. \cite{WK}.

One can ask whether the SSB approach is appropriate \cite{LS}
and whether it gives complete information \cite{LM-1}. Indeed we
will show that the assumption that the condensates are described by
Eq.\ (\ref{initialstate}) gives a broader range of physical possibilities, which are
unavailable when using Eq.\ (\ref{eq:Phase state}). The
additional effects involve phase \emph{off}-diagonal terms,
which can result in (I) violations of local realism,
i.e. violations of Bell inequalities, and (II) the occurrence of
quantum interference between macroscopically distinct states (QIMDS), as discussed by Leggett \cite{AJL}. Neither of
these effects is available in the SSB treatment. We have previously discussed \cite{LM-1}
violations of Bell inequalities with double Fock states.
Here we will show that the effect II can be detected in interferometer
experiments by the observation of ``population oscillations'',
first introduced by  Dunningham et al \cite{Dunn} within a three-condensate position interference analysis.
Theses oscillations are more robust than Bell inequality measurements,
since a few missed particles can be tolerated.

Leggett \cite{AJL} considers how one might test for QIMDS by finding coherent superpositions involving large numbers of particles (``Schr\"{o}dinger cats''). One can tell the
difference between such a pure state and a statistical mixture of
the elements of the state only by observing the off-diagonal matrix
elements between the different wave-function elements. For example,
in a state of the form $\Psi=c_{a}\psi_{a}(1,2,3,\cdots,N)+c_{b}\psi_{b}(1,2,3,\cdots,N)$
one hopes to see terms like $\left\langle \psi_{a}\left|G\right|\psi_{b}\right\rangle $
and its complex conjugate, where $G$ is an appropriate $N$-body
operator connecting the two states. As Leggett \cite{AJL} says, $\lyxmathsym{\textquotedblleft}\cdots$
what matters is that not one but a large number of elementary constituents
are behaving quite differently in the two branches.'' Here we discuss
an experiment where particles from each of two Bose condensate sources
are either deviated via a beam splitter to a side collector or proceed
to an interferometer. The measurements in the interferometer create
the two branches, and the detection in the side detectors (involving
the connecting $G$ operator) allows the observation of the off-diagonal
matrix elements of the two components. In recent years several experiments
have begun to make progress toward the goal set by Leggett, by use of
large atoms \cite{LargeAtoms}, superconductors \cite{supercon},
magnetic molecules \cite{magMole}, a quantum dot ``molecule''
\cite{quantum dot}, and photons \cite{photon}, and including a Bell
inequality violation in a Josephson phase qubit \cite{BellSuper}.

\begin{figure}[h]
\centering \includegraphics[width=2.5in]{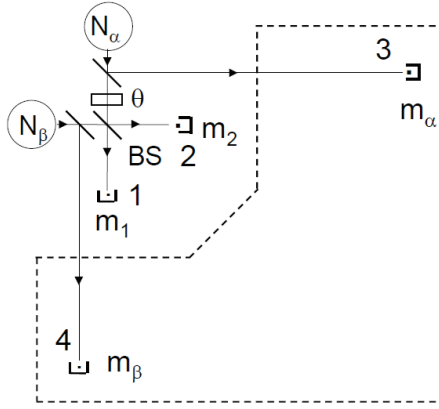}
\caption{Two source condensates, with populations $N_{\alpha}$
and $N_{\beta}$, emit particles. Some of them
reach the central beam splitter BS, followed by detectors 1 and 2 registering
$m_{1}$ and $m_{2}$ counts. The other particles are then described by a quantum superposition of macroscopically distinct states
propagating inside the region shown with a dotted line; they eventually reach counters 3 and 4, which register $m{}_{\alpha}$ and $m_{\beta}$ counts respectively. A phase shift $\theta=\pi/2$ occurs in one path.}
\label{Interferomenter}
\end{figure}

Fig. \ref{Interferomenter} shows the interferometer. The QIMDS state
is created from the double Fock state by an interference measurement at beam splitter BS, with detectors 1 and 2 giving results $m_{1},m_{2}$. The path difference between the two sources to BS is represented by angle $\theta$. Detectors 3 and 4 record $m_{\alpha}$
and $m_{\beta}$ particles respectively; although they seem to measure only the source populations, they are actually sensitive to QIMDS, as we will see.

With a single quantum particle crossing two slits, which act as sources giving rise to interference, one can measure either the interference pattern and have access to the relative phase of the sources, or from which source the particle comes; they are exclusive measurements. Here, because condensates provide many particles in the same quantum state, some of them can be used for a phase measurement, others for a source measurement.

The destruction operators $a_{1},a_{2},a_{3}$ and $a_{4}$ associated with the output modes
of the interferometer can be written in terms of the
source mode operators, $a_{\alpha}$ and $a_{\beta}$, by tracing back from the detectors to the
sources, with a factor $1/\sqrt{2}$ at each beam splitter and a phase shift of $\pi/2$ at each reflection:
\begin{eqnarray}
a_{1} & = &
\frac{1}{2}\left(e^{i\theta}a_{\alpha}+ia_{\beta}\right);\qquad
a_{2}=\frac{1}{2}\left(ie^{i\theta}a_{\alpha}+a_{\beta}\right)\nonumber
\\
a_{3} & = &
\frac{i}{\sqrt{2}}a_{\alpha};\qquad\text{              }a_{4}=\frac{i}{\sqrt{2}}a_{\beta}\label{operators}\end{eqnarray}
The probability amplitude for finding particle numbers
$\{m_{1},m_{2},m_{\alpha},m_{\beta}\}$
is:

\begin{equation}
C_{m_{1},m_{2},m_{\alpha},m_{\beta}}=\left\langle
0\left|\frac{a_{3}^{m_{\alpha}}a_{4}^{m_{\beta}}a_{1}^{m_{1}}a_{2}^{m_{2}}}{\sqrt{m_{1}!m_{2}!m_{\alpha}!m_{\beta}!}}\right|N_{\alpha}N_{\beta}\right\rangle
\label{C}\end{equation}
The double Fock state (DFS) $\left|N_{\alpha}N_{\beta}\right\rangle $
can be expanded in phase states as:
\begin{equation}
\left|N_{\alpha}N_{\beta}\right\rangle
=\sqrt{\frac{2^{N}N_{\alpha}!N_{\beta}!}{N!}}\int_{-\pi}^{\pi}\frac{d\phi}{2\pi}e^{-iN_{\beta}\phi}\left|\phi,N\right\rangle
\label{N,N}
\end{equation}
where the phase state having constant total numbers of particles is
given by Eq.\   (\ref{eq:Phase state}). These states have the property
that, for $a_{i}=v_{i\alpha}a_{\alpha}+v_{i\beta}a_{\beta}$ ($i=1,2$):
\begin{equation}
a_{i}\left|\phi,N\right\rangle
=\sqrt{\frac{N}{2}}(v_{i\alpha}+v_{i\beta}e^{i\phi})\left|\phi,N-1\right\rangle
\label{action of a}
\end{equation}
so that the state created by the interferometer is:
\begin{equation}
\left|\Gamma\right\rangle \equiv
a_{1}^{m_{1}}a_{2}^{m_{2}}\left|N_{\alpha}N_{\beta}\right\rangle
\sim\int_{-\pi}^{\pi}\frac{d\phi}{2\pi}e^{-iN_{\beta}\phi}R(\phi)\left|\phi,N-M\right\rangle
\label{eq:gamma}\end{equation}
where $M=m_{1}+m_{2}$ and:
\begin{eqnarray}
R(\phi) & = &
(e^{i\theta}+ie^{i\phi})^{m_{1}}(ie^{i\theta}+e^{i\phi})^{m_{2}}\end{eqnarray}
If we take $\theta=\pi/2$ (as we do henceforth) this takes the simple
form:
\begin{equation}
R(\phi)=(2ie^{i\phi/2})^{M}\left(\cos\frac{\phi}{2}\right)^{m_{1}}\left(\sin\frac{\phi}{2}\right)^{m_{2}}
\end{equation}
\begin{figure}[h]
\centering \includegraphics[width=2.5in]{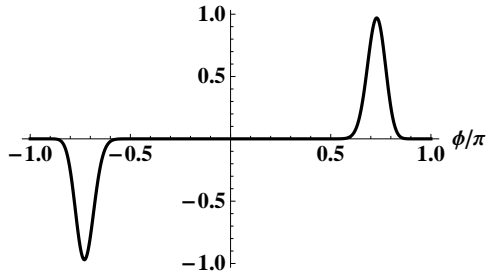}
\caption{Variations of $\hat R(\phi)$ if results $m_{1}=17$ and
$m_{2}=83$ are obtained.
The peaks are at $\phi_0=\pm0.73\pi$ (the phase choice $\theta=\pi/2$ gives symmetrical peaks about zero).
The relative sign of the two peaks is $(-1)^{m_{2}}$. For large numbers of particles, the
measurement produces a coherent superposition of macroscopically distinct states (``Schr\"{o}dinger cat'').}
\label{R(phi)}
\end{figure}
Fig. \ref{R(phi)} shows $\hat{R}(\phi)=R(\phi)\times(2ie^{i\phi/2})^{-M}$, which has two peaks at
$\pm \phi_0=\pm2\arctan \sqrt{m_{2}/{m_{1}}}$. This is not surprising: classically, the ratio of the
intensities in the output arms of the interferometer determines the absolute value or the phase difference between the two input beams, but
not its sign. Separating the negative and positive contribution of $\phi$ provides:
\begin{equation}
\left|\Gamma\right\rangle =\left|\psi_{+}\right\rangle
+(-1)^{m_{2}}\left|\psi_{-}\right\rangle \label{eq:DPhS}\end{equation}

Let us begin with a qualitative calculation. We assume that $M$ is large, so that the peaks are sharp and:
\begin{equation}
\left|\psi_{\pm}\right\rangle \sim e^{\mp
i(N_{\beta}-M/2)\phi_0}\left|\pm\phi_0,N-M\right\rangle\label{approxim}
\end{equation}
These two wave-function branches are orthogonal for large $M$ for
any $\phi_0$ not too near zero; and they are macroscopic as long
as $N-M$ is large.

Showing QIMDS requires making a measurement that is sensitive to
the interference between the two components; this is the role of the side-detectors
shown in Fig. \ref{Interferomenter}. Because
$a_{3}^{m_{\alpha}}a_{4}^{m_{\beta}}\left|\pm\phi_0,N-M\right\rangle
\sim e^{\pm im_{\beta}\phi_0}\left|0\right\rangle $,
the probability of getting the set
$\{m_{1},m_{2},m_{\alpha},m_{\beta}\}$
becomes:
\begin{equation}
P(m_{1},m_{2},m_{\alpha},m_{\beta})  \sim
1+(-1)^{m_{2}}\cos\left[\left(m_{\alpha}-m_{\beta}\right)\phi_0\right]\label{eq:QualitativePO}
\end{equation}
(if $N_{\alpha}=N_{\beta}$), where the cosine terms arises from the
sum of the two cross terms
$\left\langle \psi_{\pm}\left|a_{3}^{\dagger
m_{\alpha}}a_{4}^{\dagger
m_{\beta}}a_{3}^{m_{\alpha}}a_{4}^{m_{\beta}}\right|\psi_{\mp}\right\rangle
$. Now, if one does the interferometer experiment
for fixed source numbers, say, $N_{\alpha}=N_{\beta}$, and considers
only those experiments having the same $m_{1},\text{ }m_{2}$, then
the interference between the two elements will show up in a cosine
variation of probability with $m_{\alpha}.$ We call this effect ``population oscillations''; it was already discussed in Ref. \cite{Dunn} for three-condensate
experiments.

These oscillations are beyond SSB since they disappear if one starts from either (\ref{eq:BrokenSym}) or (\ref{eq:Phase state}). With a phase state of phase $\chi$ for instance, the action of the destruction operators $a_{1,2}$ on this state introduces $\chi$ instead of an integration variable $\phi$ into (\ref{action of a}); this leads essentially to (\ref{eq:gamma}) without the $\phi$ integral. No interference effect between two phase peaks occurs and the probability is proportional to $|R(\chi)|^2$. One gets a $m_{\alpha}, m_{\beta}$ dependence of the probability that is proportional to a simple binomial distribution $(N-M)!/m_{\alpha}!m_{\beta}! $, without any oscillation. Actually the angle $\chi$ plays no role at all in this dependence, which is natural since detectors 3 and 4 do not see an interference effect between two beams; they just measure the intensities of two independent sources after a beam splitter at their output.

A more accurate calculation is now presented. Operating on Eq.\   (\ref{eq:gamma}) with
$a_{3}^{m_{\alpha}}a_{4}^{m_{\beta}}$,
and forming the probability
introduces another angle $\phi^{\prime}$,
so that the probability for finding the set
$\{m_{1},m_{2},m_{\alpha},m_{\beta}\}$
takes the form:
\begin{widetext}
\begin{equation}
P(m_{1},m_{2},m_{\alpha},m_{\beta})=\frac{N_{\alpha}!N_{\beta}!}{m_{1}!m_{2}!m_{\alpha}!m_{\beta}!}\frac{1}{2^{N+M}}\int_{-\pi}^{\pi}\frac{d\phi^{\prime}}{2\pi}\int_{-\pi}^{\pi}\frac{d\phi}{2\pi}e^{-i(N_{\alpha}-m_{\alpha})(\phi-\phi^{\prime})}R^{*}(\phi^{\prime})R(\phi)\end{equation}
We note that one phase branch peak occurs for $-\pi\le\phi\le0$
and the other for $0\leq\phi\le\pi$, so that the overlap between
different
branches occurs for $\phi^{\prime}\ne\phi.$ A change of variables
to $\Lambda=(\phi-\phi^{\prime})/2$ and
$\lambda=\pi/2-\theta+(\phi+\phi^{\prime})/2$
leads to:
\begin{eqnarray}
P(m_{1},m_{2},m_{\alpha},m_{\beta}) & = &
\frac{N_{\alpha}!N_{\beta}!}{m_{1}!m_{2}!m_{\alpha}!m_{\beta}!2^{N}}\int_{-\pi}^{\pi}\frac{d\Lambda}{2\pi}\cos\left[\left(N_{\alpha}-m_{\alpha}-N_{\beta}+m_{\beta}\right)\Lambda\right]\int_{-\pi}^{\pi}\frac{d\lambda}{2\pi}\nonumber
\\
 &  &
\times\left[\cos\Lambda+\cos\lambda\right]^{m_{1}}\left[\cos\Lambda-\cos\lambda\right]^{m_{2}}\label{eq4}\end{eqnarray}
\end{widetext}
When $\theta=\pi/2$ the ``classical phase angle'' $\lambda$ is half the sum of $\phi$ and
$\phi^{\prime}$. The expresssion also contains another angle, $\Lambda$, which we call
the ``quantum angle'' - in Ref. \cite{LM-1} it appeared as a consequence of a conservation rule,
but here we introduce it to characterize quantum interference effects between different values of the phase.

\begin{figure}[h]
\centering \includegraphics[width=2.5in]{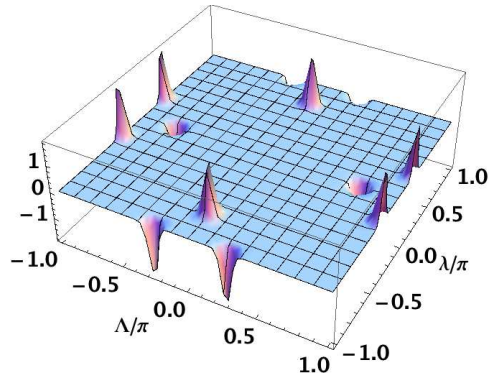}

\caption{Plot of $F(\Lambda,\lambda)$ as a function of $\Lambda$ and
$\lambda$
for $m_{1}=17$ and $m_{2}=83$. The peaks along $\Lambda=0$ and
$\pm\pi$ correspond to phase diagonal matrix elements,
while the negative depressions, having $\Lambda\ne0,$ correspond
to off-diagonal matrix elements between two macroscopic phases (QIMDS). If $m_{1}$ and $m_{2}$ are even, the negative depressions become positive peaks. }
\label{LamlamPlot}
\end{figure}

To examine the behavior of the probability, we plot the quantity:
\begin{equation}
F(\Lambda,\lambda)=\left[\cos\Lambda+\cos\lambda\right]^{m_{1}}\left[\cos\Lambda-\cos\lambda\right]^{M-m_{1}}\label{F}\end{equation}
 We take $N_{\alpha}=N_{\beta}=M=100$ in our examples here.
$F(\Lambda,\lambda)$
has multiple peaks as shown in Fig. \ref{LamlamPlot} for $m_{1}=17$.
The extrema are easily shown to occur at:
\begin{eqnarray}
\Lambda & = &
0\quad\mathrm{and}\quad\lambda=\pm2\arctan\left(\sqrt{m_{2}/m_{1}}\right);\nonumber
\\
\Lambda & = & \pm \pi\quad
\mathrm{and}\quad\lambda=\pm2\arctan\left(\sqrt{m_{1}/m_{2}}\right);\nonumber
\\
\Lambda & = &
\pm2\arctan\left(\sqrt{m_{2}/m_{1}}\right)\quad\mathrm{and}\quad\lambda=0\;;\nonumber
\\
\Lambda & = &
\pm2\arctan\left(\sqrt{m_{1}/m_{2}}\right)\quad\mathrm{and}\quad\lambda=\pm\pi\end{eqnarray}
For $m_{1}=17,$ $m_{2}=83$ we have peaks
$(\Lambda,\lambda)=(0,\pm2.29),$
$(\pi,\pm0.85)$ and depressions at
$(\Lambda,\lambda)=(\pm2.29,0),\text{ }(\pm0.85,\pm\pi)$,
where $\pi-2.29=0.85.$ These extrema are precisely at the positions given by the elements of the
density matrix associated with state (\ref{approxim}).

The peaks along $\Lambda=0$ (and $\pm\pi$) correspond
to phase-diagonal matrix elements which, in (\ref{eq4}), introduce
the usual probabilities $\left[1\pm\cos\lambda\right]$ associated with an interferometer,
averaged oven a random phase $\lambda$. The extrema centered at $\Lambda\ne0$ are phase off-diagonal, and directly indicate QIMDS since here the phase state is macroscopic.
When $\Lambda$ does not vanish, probabilities become quasi-probabilities $\left[\cos\Lambda\pm\cos\lambda\right]$,
which may be negative.

\begin{figure}[h]
\centering \includegraphics[width=2.5in]{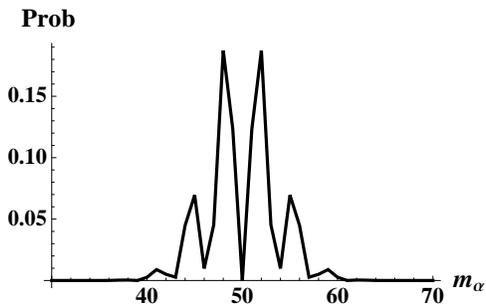}
\caption{Plot of $P(m_{1},m_{\alpha})$ given by Eq.\ (\ref{eq4})
versus $m_{\alpha}$
for $N_{\alpha}=N_{\beta}=M=100,$ $m_{1}=17$ and $m_{2}=83$. If $m_{2}$ is even, the central dip
is replaced by a peak.}
\label{DLam&PO}
\end{figure}

Now, in (\ref{eq4}), the $m_{\alpha}$ and $m_{\beta}$ dependence is
given by the cosine Fourier transform of a function obtained by integrating $F(\Lambda,\lambda)$ in Eq.(\ref{F}) over
$\lambda$. Because $F(\Lambda,\lambda)$
has multiple peaks, the final probability contains oscillations as a function
of $m_{\alpha}$ as shown in Fig. \ref{DLam&PO} - if we replace
the peaks in $F(\Lambda,\lambda)$ with $\delta$-functions we recover
exactly Eq. (\ref{eq:QualitativePO}). By contrast, within SSB the result is equivalent to Eq. (\ref{eq4}), but with $\Lambda$ set to zero,
which cancels
the contribution of the off-diagonal peaks. The probabilities then become
smooth functions of $m_{\alpha}$, with no dips or peaks; population oscillations disappear.

If we did not count $m_{\alpha}$ and $m_{\beta}$ but summed
over these variables with a given sum $m_{\alpha}+m_{\beta}=N-M$,
we would get a factor $\left(\cos\Lambda\right)^{N-M}$, strongly peaked
at $\Lambda=0$ if $N-M$ is large. The probability
of finding the result set $\{m_{1},m_{2}\}$ would then be
$P(m_{1},m_{2})\sim\int_{-\pi}^{\pi}\frac{d\lambda}{2\pi}\left[1+\cos\lambda\right]^{m_{1}}\left[1-\cos\lambda\right]^{m_{2}}$.
This still has two peaks in the integrand, which arise since
the interferometer cannot discriminate between opposite relative phases.
But what is now obtained is a statistical mixture of these two phases, without
any population oscillation; the situation is analogous
to that described by Eq.\   (\ref{classical}).

The analysis of Bell violations in Ref.\ \cite{LM-1}
shows that one single missed particle cancels the violation.
The population oscillations have no special relation to locality, and they are more robust. We have shown that, by proper
selection of $m_1$ and $m_2$,
one can preserve a small central with as many as 5 particles lost.

In conclusion, two kinds of interference
effects occur.  One produces the fringes seen in the MIT experiments in the
merging of two Bose condensates.  This effect can be explained by using
SSB and either Eq.\
(\ref{eq:BrokenSym}) or Eq.\ (\ref{eq:Phase state}). But the approach using a double Fock
state preserves a second
interference effect: the macroscopic quantum interference that
involves the off-diagonal elements corresponding to $\Lambda\neq0$,
and leads to QIMDS that are observable via ``population oscillations''.

\ Laboratoire Kastler Brossel
is \textquotedblleft UMR 8552 du CNRS, de l'ENS, et de
l'Universit\'{e}
Pierre et Marie Curie\textquotedblright .

\end{document}